# Nonlinear vectorial prediction with neural nets[1]

Marcos Faúndez-Zanuy

Escola Universitària Politècnica de Mataró
Universitat Politècnica de Catalunya (UPC)
Avda. Puig i Cadafalch 101-111, E-08303 Mataró (BARCELONA) SPAIN
`faundez@eupmt.es`

**Abstract.** In this paper we propose a nonlinear vectorial prediction scheme based on a Multi Layer Perceptron. This system is applied to speech coding in an ADPCM backward scheme. In addition a procedure to obtain a vectorial quantizer is given, in order to achieve a fully vectorial speech encoder. We also present several results with the proposed system

## 1 Introduction

Most of the speech coders use some kind of prediction. The most popular one is the scalar linear prediction, but several papers have shown that a nonlinear predictor can outperform the classical LPC linear prediction scheme. On the other hand, vectorial prediction schemes have also been proposed. In this paper we propose and study a vectorial nonlinear prediction scheme based on a MLP neural net. We apply the nonlinear predictor inside an ADPCM scheme for speech coding. This predictor replaces the LPC predictor in order to obtain an ADPCM speech encoder scheme with nonlinear prediction. Figure 1 summarizes the scheme. If the quantizer is a vectorial quantizer, this scheme is also known as Predictive Vector Quantization [1].

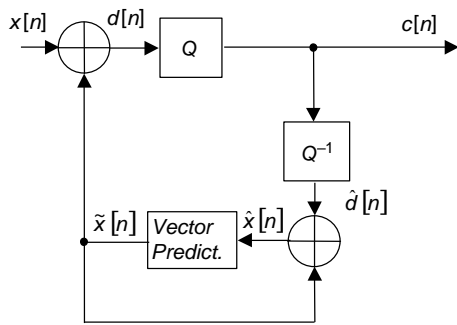

**Fig. 1.** Vectorial ADPCM scheme.

[1] This work has been supported by the CICYT TIC2000-1669-C04-02

---

Classical ADPCM block-adaptive waveform coders compute the predictor coefficients in one of two ways:

a) backward adaptation: The coefficients are computed over the previous frame. Thus, it is not needed to transmit the coefficients of the predictor, because the receiver has already decoded the previous frame and can obtain the same set of coefficients.

b) forward adaptation: The coefficients are computed over the same frame to be encoded. Thus, the coefficients must be quantized and transmitted to the receiver. In [2] we found that the SEGSNR of forward schemes with unquantized coefficients is similar to the classical LPC approach using one quantization bit less per sample. On the other hand with this scheme the mismatch between training and testing phases is smaller than in the previous case, so the training procedure is not as critical as in backward schemes, and the SEGSNR are greater.

The SEGSNR is computed with the expression: $SNRSEG = \frac{1}{K}\sum_{j=1}^{K} SNR_j$ , where $SNR_j$ is the signal to noise ratio (dB) of frame $j$ : $SNR = \frac{E\{x^2[n]\}}{E\{e^2[n]\}}$, and $K$ is the number of frames of the encoded file.

In [3] we found that the quantization of the predictor coefficients (forward scheme) is not a trivial question. In [4] we proposed another training approach of the neural net, in order to improve the performance of the backward scheme. The goal is to obtain similar results with backward ADPCM nonlinear prediction and forward (unquantized coefficients) schemes. We will work with the backward scheme. Figure 2 shows the network architecture for scalar and prediction.

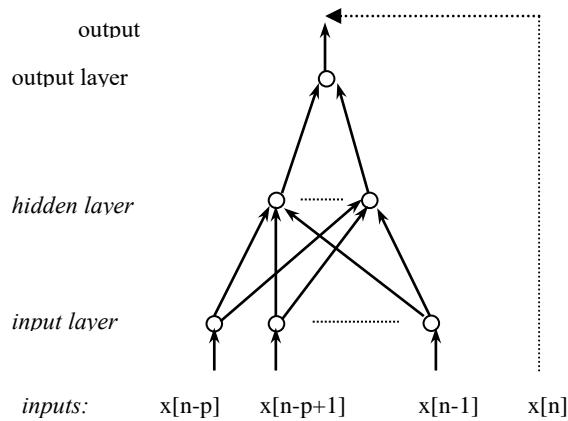

**Fig. 2.** Neural net training for scalar prediction

In our previous work we fixed the structure of the neural net to 10 inputs, 2 neurons in the hidden layer, and one output. The selected training algorithm was the Levenberg-Marquardt, that computes the approximate Hessian matrix, because it is faster and achieves better results than the classical backpropagation algorithm. We also apply a multi-start algorithm with five random initializations for each neural net. In [4] we studied the following training schemes:
a) The use of regularization.
b) Early stopping with validation.
c) Bayesian regularization.
d) Validation and bayesian regularization.
e) Committee of neural nets [5].
The combination between Bayesian regularization with a committee of neural nets increased the SEGSNR up to 1.2 dB over the MLP trained with the Levenberg-Marquardt algorithm, and decreases the variance of the SEGSNR between frames.

### 1.1 Conditions of the experiments

The experimental results have been obtained with an ADPCM speech coder with an adaptive scalar quantizer based on multipliers [7]. The number of quantization bits is variable between Nq=2 and Nq=5, that correspond to 16kbps and 40kbps (the sampling rate of the speech signal is 8kHz). We have encoded eight sentences uttered by eight different speakers (4 males and 4 females). The neural net has been trained with the Levenberg-Marquardt algorithm combined with bayesian regularization [7-8]

## 2. Vectorial prediction

In this paper we propose a different scheme, that consists on a vectorial neural net prediction.
Previously to the vectorial prediction scheme we will study the classical scalar prediction scheme, using a neural net training with hints. The proposed scheme consists on the neural net architecture shown on figure 3, where the second output is only used for training purposes. That is, x[n+1] is used as a hint for a second neural net output during training, but this output is ignored during the speech coding process.
Table 1 compares the obtained results for one and two outputs using the best approach found in [4], that consists on bayesian regularization with a committe of 5 neural nets for the prediction process.
Table 1 shows the obtained segmental signal to noise ratio SEGSNR and the standard deviation of the SEGSNR ($\sigma$).
The results of table 1 are similar to the best results that we obtained in [4] (using the scheme of figure 1) for the best situation. Table 2 reproduces these results for comparison purposes.

The next experiment consists on an ADPCM speech coding with vectorial prediction and scalar adaptive quantization (same quantizer used in the previous experiments). Table 3 shows the obtained results with a prediction of two samples simultaneously.

**Table 1.** SEGSNR for several ADPCM schemes with the nnet of figure 2 and scalar prediction.

| Epoch | combination | Nq=2 | | Nq=3 | | Nq=4 | | Nq=5 | |
|---|---|---|---|---|---|---|---|---|---|
| | | SEGSNR | $\sigma$ | SEGSNR | $\sigma$ | SEGSNR | $\sigma$ | SEGSNR | $\sigma$ |
| 6 | Mean | 14.2 | 5.1 | 20.4 | 5.8 | 25.6 | 6.3 | 30.4 | 6.7 |
| 50 | mean | 13.9 | 5.5 | 20.8 | 6.0 | 26 | 6.6 | 30.9 | 6.9 |
| 6 | median | 14.7 | 5.1 | 20.8 | 6.0 | 25.7 | 6.4 | 30.4 | 6.7 |
| 50 | median | 14.0 | 5.6 | 21 | 6.1 | 26.2 | 6.7 | 31.1 | 7.0 |

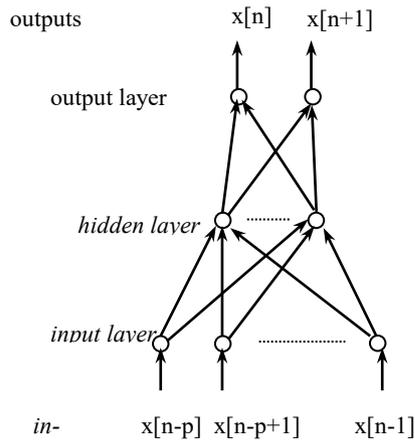

**Fig. 3.** Neural net training for vectorial prediction

**Table 2.** SEGSNR for several ADPCM schemes with the nnet of figure 1

| Epoch | combination | Nq=2 | | Nq=3 | | Nq=4 | | Nq=5 | |
|---|---|---|---|---|---|---|---|---|---|
| | | SEGSNR | $\sigma$ | SEGSNR | $\sigma$ | SEGSNR | $\sigma$ | SEGSNR | $\sigma$ |
| 6 | Mean | 14.5 | 4.9 | 20.5 | 5.9 | 25.6 | 6.5 | 30.5 | 6.8 |
| 50 | mean | 14.6 | 5.6 | 21.3 | 6.4 | 26.5 | 6.8 | 31.4 | 7.2 |
| 6 | median | 14.9 | 5.2 | 21 | 6 | 25.9 | 6.5 | 30.7 | 6.9 |
| 50 | median | 14.3 | 5.5 | 21.1 | 6.2 | 26.4 | 6.8 | 31.2 | 7.1 |

Table 3 shows that the behavior of the vectorial prediction scheme is worst than the scalar scheme. We believe that this is due to the quantizer, because the experiments of table 2 show a good performance with the same neural net scheme. It is interesting to observe that the vectorial scheme is more suitable with a vectorial quantizer. In order to check this assert we have plot the vectorial prediction errors on figure 4. It is interesting to observe that the vectors to be quantized do not lie in the whole space. Most of the vectors are around $y=x$ that means that the error vector presents the following property: $\vec{e}[n] = (e_{1n}, e_{2n})$, $e_{1n} \cong e_{2n}$. Thus, a vectorial quantizer can exploit this redundancy and improve the results.

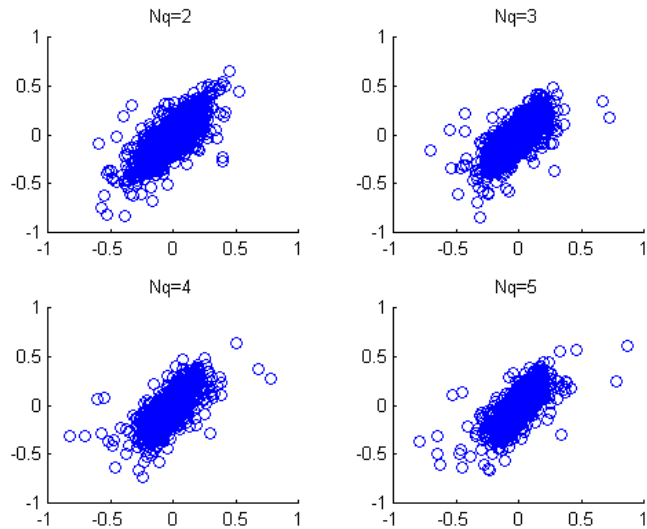

**Fig. 4.** Vectorial prediction errors for several quantization bits

**Table 3.** SEGSNR for several ADPCM schemes with the nnet of figure 2 and vectorial pred.

| Epoch | combination | Nq=2 | | Nq=3 | | Nq=4 | | Nq=5 | |
|---|---|---|---|---|---|---|---|---|---|
| | | SEGSNR | σ | SEGSNR | σ | SEGSNR | σ | SEGSNR | σ |
| 6 | Mean | 12.58 | 4.4 | 18.3 | 4.9 | 23.0 | 5.3 | 27.5 | 5.4 |
| 50 | mean | 12.9 | 5 | 18.9 | 5.4 | 23.7 | 5.5 | 28.3 | 5.8 |
| 6 | median | 12.9 | 4.5 | 18.6 | 5.2 | 23.2 | 5.3 | 27.7 | 5.5 |
| 50 | median | 12.9 | 4.8 | 18.8 | 5.3 | 23.6 | 5.5 | 28.1 | 5.7 |

**Non-linear predictive vector quantization**
A special case of vector quantization is known as predictive vector quantization (PVQ) [1]. Basically, PVQ consists on an ADPCM scheme with a vector predictor and a vectorial quantizer. Obviously, if the vector predictor is nonlinear the system is a NL-PVQ scheme. We propose to use the NL-vector predictor described in previous sections, and a vectorial quantizer. In order to design the vectorial quantizer we use the residual signal of the vectorial prediction and the scalar quantizer based on multipliers applied as many times as the dimension of the vectors. Thus, we have used the residual errors of the first speaker (with Nq=3) and the generalized Lloyd algorithm [1] in order to create codebooks ranging from 4 to 8 bits. The initial codebook has been obtained with the random method.

It is interesting to observe that the optimization procedure must be a closed loop algorithm, because there are interactions between the predictor and the quantizer. Thus, the system can be improved computing again the residual errors with the actual quantization scheme

Figure 5 shows the obtained codebooks and table 4 the SEGSNR for several quantization bits per sample (this value is obtained by the ratio between the number of bits of the vectorial quantizer and the dimension of the vectors). The committee of neural nets is obtained with the *median{}* function and 50 epochs. It can be seen that the Vector Quantizer results of table 4 outperform the scalar quantizer of table 3 in nearly 2 dB for Nq=4.

**Table 4.** SEGSNR for NL-PVQ.

| Nq=2 | | Nq=2.5 | | Nq=3 | | Nq=3.5 | | Nq=4 | |
|---|---|---|---|---|---|---|---|---|---|
| SEGSNR | $\sigma$ | SEGSNR | $\sigma$ | SEGSNR | $\sigma$ | SEGSNR | $\sigma$ | SEGSNR | $\sigma$ |
| 10.4 | 8.3 | 15.8 | 6.9 | 19 | 5.9 | 21.4 | 6.1 | 25 | 5.7 |

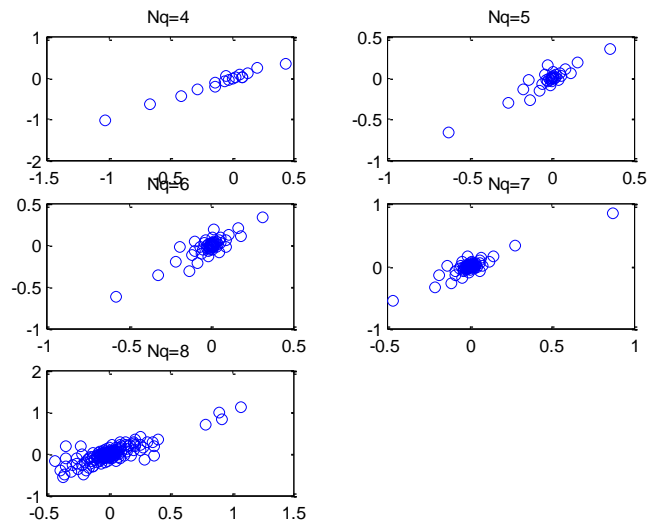

**Fig. 5.** Codebooks with a random initialization and generalized Lloyd iteration.

**Scalar linear prediction**

The AR modeling of order P is given by the following relation:

$$x[n] = \sum_{i=1}^{P} a_i x[n-i] + e[n]$$

where $\{a_i\}_{i=1,\ldots P}$ are the scalar prediction coefficients. Their value is usually obtained with the levinson-durbin recursion [1].

**Vectorial linear prediction**

The AR-vector modeling of order P is given by the following relation:

$$\vec{x}[n] = \sum_{i=1}^{P} A_i \vec{x}[n-i] + \vec{e}[n]$$ where $\{A_i\}_{i=1,\ldots P}$ are the $m \times m$ matrices equivalent to the prediction coefficients of the classical scalar predictor, and *m* is the dimension of the vectors. The prediction matrices can be estimated using the Levinson-Whittle-Robinson algorithm, that has been previously applied to speaker verification in [5]. The algorithm is the following:

Computation of the correlation matrices $R_i$:

$$R_i = \sum_{n=0}^{N-i-1} \vec{x}[n+i](\vec{x}[n])^T$$

where $(\vec{x}[n])^T$ means the transpose of the vector $\vec{x}[n]$.

for i=0:P-1,

$$F_i = \sum_{j=0}^{i} A1_j^i R_{i+1-j}$$

$$K1_i = -F_i (D2_i)^{-1}$$

$$K2_i = -F_i^T (D1_i)^{-1}$$

$$D1_{i+1} = (I - K1_i K2_i) D1_i$$

$$D2_{i+1} = (I - K2_i K1_i) D2_i$$

$$A1_0^{i+1} = A1_0^i$$

$$A2_0^{i+1} = K2_i A1_0^i$$

$$A1_{i+1}^{i+1} = K1_i A2_i^i$$

$$A2_{i+1}^{i+1} = A2_i^i$$

for j=1:i,

$$A1_j^{i+1} = A1_j^i + K1_i A2_{j-1}^i$$

$$A2^{i+1}_j = A2^i_{j-1} + K2_i A1^i_j$$

      end
end
for i=0:P,
$$A_i = A2^P_i$$
end

## Conclusions and future work

A PVQ system implies a good predictor and a good quantizer. In this paper we have first proposed a MLP training with output hints, in order to check that the proposed scheme for vectorial prediction is suitable. The next step has been the evaluation of the system with scalar and vectorial quantizers, and we have found that the SEGSNR is smaller than the achieved one with the scalar system. we believe that this is because of the quantizer. Thus, the next step will be the evaluation on the PVQ system proposed in this paper with an improved quantizer in two ways:
   a) The VQ will be computed with a better algorithm, like the Linde-Buzzo-Gray (LBG) one. Looking at figure 5 it can be seen that the centroids are not well distributed along the vectorial space.
   b) The VQ is computed in an open loop design, and it is used in a closed loop system. An iterative algorithm [10] must be applied in order to jointly optimize the predictor and the quantizer.

Future work will include the study of vectorial prediction of order higher than 2. In addition, the PVQ scheme will be the first step towards a CELP speech coder.

## References

[1] A. Gersho & R. M. Gray "Vector Quantization and signal compression". Ed. Kluwer 1992.
[2] M. Faundez-Zanuy, F. Vallverdu and E. Monte, "Nonlinear prediction with neural nets in ADPCM," Proceedings of the 1998 IEEE International Conference on Acoustics, Speech and Signal Processing, ICASSP '98 (Cat. No.98CH36181), 1998, pp. 345-348 vol.1, doi: 10.1109/ICASSP.1998.674438
[3] O. Oliva, M. Faúndez "A comparative study of several ADPCM schemes with linear and nonlinear prediction" EUROSPEECH'99 , Budapest, Vol. 3, pp.1467-1470.
[4] M. Faúndez-Zanuy, "Nonlinear predictive models computation in ADPCM schemes". Vol. II, pp 813-816. EUSIPCO 2000, Tampere.
[5] C. Montacié & J. L. Le Floch, "Discriminant AR-Vector models for free-text speaker verification", pp.161-164, Eurospeech 1993.
[6] S. Haykin, "neural nets. A comprehensive foundation", 2on edition. Ed. Prentice Hall 1999.

[7] N. S. Jayant and P. Noll "Digital Coding of Waveforms". Ed. Prentice Hall 1984.
[8] D. J. C. Mackay "Bayesian interpolation", Neural computation , Vol.4, Nº 3, pp.415-447, 1992.
[9] F. D. Foresee and M. T. Hagan, "Gauss-Newton approximation to Bayesian regularization", proceedings of the 1997 International Joint Conference on Neural Networks, pp.1930-1935, 1997.
[10] V. Cuperman, A. Gersho "Vector predictive coding of speech at 16 kbits/s". IEEE trans. on Commun. vol. COM-33, pp.685-696, July 1985.